# Does the price of strategic commodities respond to U.S. Partisan Conflict?

Yong Jiang[a], Yi-Shuai Ren[b,c,d,e]*, Chao-Qun Ma[c,d,f], Jiang-Long Liu[d,f], Basil Sharp[e,g]

[a] *School of Finance, Nanjing Audit University, Nanjing 211815, China*
[b] *School of Public Administration, Hunan University, Changsha 410082, China*
[c] *Center for Resource and Environmental Management, Hunan University, Changsha 410082, China*
[d] *Research Institute of Digital Society and BlockChain, Hunan University, China*
[e] *The Energy Centre, University of Auckland, 12 Grafton Rd, Auckland 1010, New Zealand*
[f] *Business School, Hunan University, Changsha 410082, China*
[g] *The University of Auckland, Auckland 1142, New Zealand*

**Abstract**

A noteworthy feature of U.S. politics in recent years is serious partisan conflict, which has led to intensifying polarization and exacerbating high policy uncertainty. The US is a significant player in oil and gold markets. Oil and gold also form the basis of important strategic reserves in the US. We investigate whether U.S. partisan conflict affects the returns and price volatility of oil and gold using a parametric test of Granger causality in quantiles. The empirical results suggest that U.S. partisan conflict has an effect on the returns of oil and gold, and the effects are concentrated at the tail of the conditional distribution of returns. More specifically, the partisan conflict mainly affects oil returns when the crude oil market is in a bearish state (lower quantiles). By contrast, partisan conflict matters for gold returns only when the gold market is in a bullish scenario (higher quantiles). In addition, for the volatility of oil and gold, the predictability of partisan conflict index virtually covers the entire distribution of volatility.

**Keywords**: U.S. partisan conflict; Granger causality in quantiles; oil prices; gold prices

# 1. Introduction

Faced with international turbulence and a slowdown in global economic recovery, oil and gold reserves play an important positive role in maintaining economic stability. Oil and gold are deemed to have properties similar to of ordinary commodities, and their prices are mainly influenced by factors underpinning supply and demand (Cai et al., 2001; Tully and Lucey, 2007; Kilian, 2009; Gallo et al., 2010; Coleman, 2012; Kim and Vera, 2018). However, oil and gold are also strategic commodities used for hedging against inflation, portfolio diversification, and a safe haven in times of political turbulence and severe market turmoil (Baur and McDermott, 2010; Aye et al., 2016). Therefore, improving the understanding of drivers of the price of oil and gold is a longstanding research objective. A large number of studies find evidence that oil and gold prices are influenced by non-fundamentals such as the economic policy uncertainty (EPU), investor attention and political risk (Coleman, 2012; Chen et al., 2016; Wang and Sun, 2017; Yao et al., 2017; Uddin et al., 2018). For example, Jones and Sackley (2016) incorporate the U.S. EPU index into a gold-pricing model and find that gold prices are positively related to EPU.

As the largest consumer, and second-largest importer, of crude oil in the world, the U.S. occupies an important place both in the crude oil market (BP Statistical Review of World Energy, 2018). The U.S. consumed 913.3 million tonnes oil equivalent in 2017, accounting for 19.8% of the world's consumption of crude oil. Since 2009, with the progress in shale oil recovery techniques, U.S. crude oil production has gradually increased to become the world's second-largest oil producer, just after Saudi Arabia. In 2017, it produced 571 million tons, accounting for a global share of 13%. In addition, the U.S. has the largest gold reserves in the world, reaching 8133.5 tons in 2017, accounting for 72.65% of its foreign exchange reserves. Therefore, given the position of the U.S. in global commodity markets, we cannot ignore the role that the U.S. factors play in the pricing of international commodities (Kang and Ratti, 2013a; Kang and Ratti, 2013b; Balcilar et al., 2017a).

In recent years, U.S. politics have been characterized by a high degree of partisan conflict. The combination of increasing polarization with the divided government may affect the decisions of investors in financial markets. US partisan conflict affects private investment (Azzimonti, 2018) or has a critical impact on the U.S. stock market (Gupta et al., 2018a; Gupta et al., 2018b). Fig.1 displays the dynamic of oil prices, gold prices and U.S. partisan conflict index[1]. As can be seen, the partisan conflict index tends to increase with the forthcoming election and during debates over such contentious policies as the debt ceiling and health-care reform. The rise in partisan conflict was accelerated during the Great Recession and reached a peak in 2013 at the time of the government shutdown. After the financial crisis in 2008, the US partisan conflict index had a significant upward trend, and the prices of crude oil and gold have a corresponding upward trend. During the Trump-Clinton election of 2016, the partisan conflict index up to a peak. Spearman's rank correlation coefficient of the partisan conflict index and the crude oil price is 0.3, with the price of gold, it is 0.68. These results suggest that there exists a link between the partisan conflict index and the price of crude oil and gold. However, it is unclear as to whether the effect of the U.S. partisan conflict index is linear or nonlinear. Furthermore, is the effect of the partisan conflict index homogeneous under different market conditions? Answers to these questions are significant for managing risk, hedging and making investment decisions.

To address these issues, we use monthly data on the oil and gold prices and the U.S. partisan conflict indices from January 1981 to October 2017. We employ a parametric test of Granger causality in quantiles proposed by Troster (2018) to study whether the U.S. partisan conflict can influence the returns and volatility for oil and gold market. We find that there is no significant evidence to support a causal link between the U.S. partisan conflict and returns for oil and gold at the median of the conditional distribution. The explanatory power of the U.S. partisan conflict index on oil and gold returns tends to cluster around at the tails of the conditional distribution. More specifically, the partisan conflict has strong predictive power on the oil returns when the crude oil market is bearish. However, the impact of the partisan conflict on oil returns is not significant in a bullish market. By contrast, partisan conflict impacts gold returns only when the gold market is bullish. At lower quantiles, the partisan conflict index has limited predictive power, with the 0.2 quantiles being an exception. Generally, the U.S. partisan conflict significantly affects the volatility of oil and gold prices over the entire conditional distribution, namely, at various phases of the oil and gold market.

---

[1] We use US refiner's acquisition cost of crude oil as the measure of the global crude oil prices and gold price is obtained from London Bullion Market. U.S. partisan conflict index (PCI) was developed by Azzimonti (2018), which tracks the degree of political disagreement among U.S. politicians at the federal level by measuring the frequency of newspaper articles reporting disagreement in a given month. Higher index values indicate greater conflict among the political parties, congress, and the president.

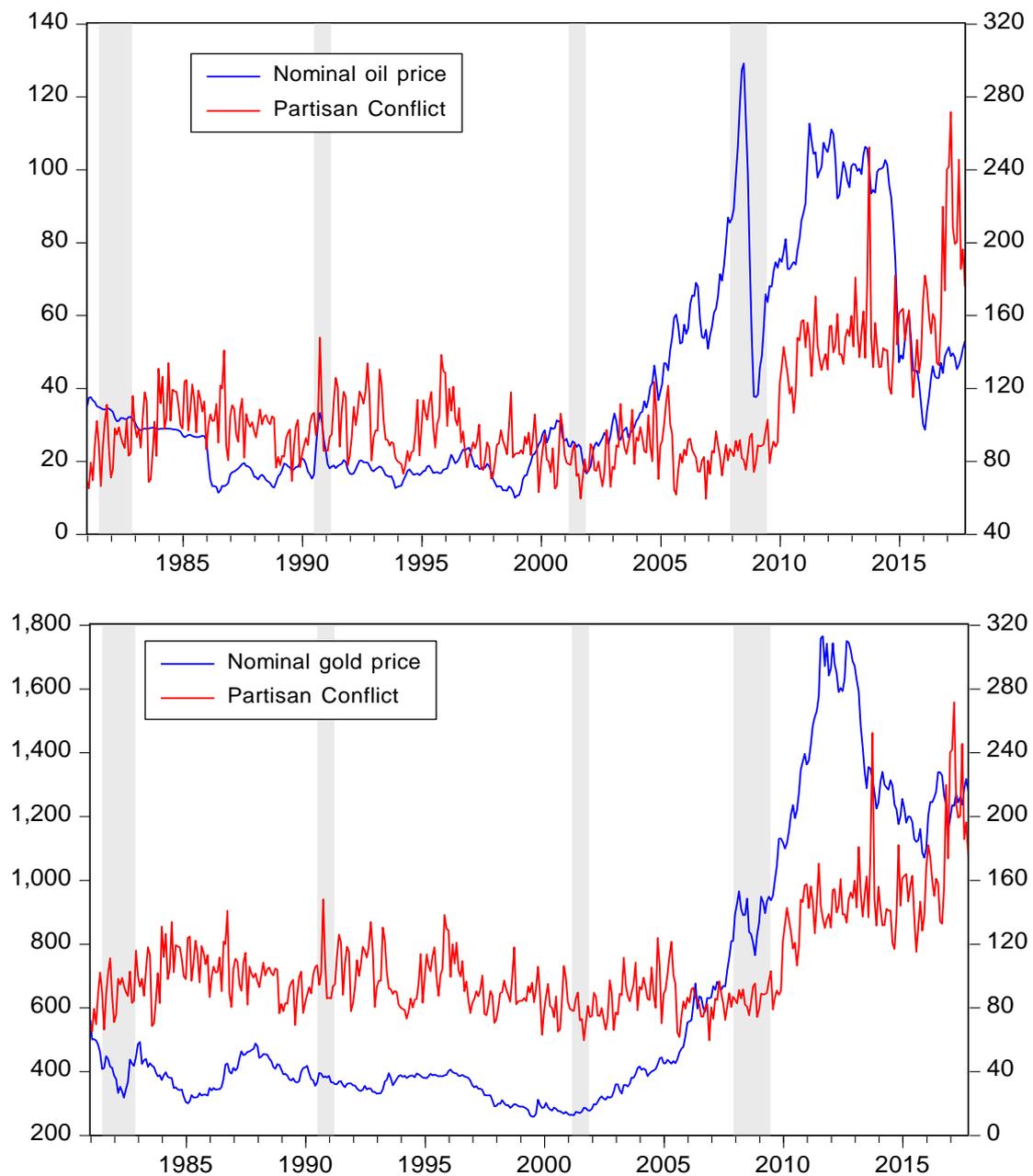

**Fig. 1** Oil and gold prices and U.S. partisan conflict index[2]

This paper makes several contributions to the literature as follows. First, this paper examines the predictive effect of U.S. partisan conflict on the returns and volatility for oil and gold. Although some studies have considered the effect of non-fundamentals such as the EPU and investor sentiment, little research has examined whether the U.S. partisan conflicts have an impact on returns and volatility for oil and gold. By doing this, we provide a new predictor that can easily be observed s to track the dynamic characteristics of crude oil and gold prices.

---

[2] Right axis is U.S. partisan conflict index. The shaded regions represent NBER recessions.

Second, we focus on the volatility estimation of oil and gold under the framework of the generalized autoregressive conditional heteroscedastic (GARCH) models and the stochastic volatility (SV) models. Unlike previous papers that characterize volatility by using time-varying volatility models directly such as GARCH and SV models (Hammoudeh and Yuan, 2008; Ewing and Malik, 2016), we compare several GARCH and SV models in a formal Bayesian model comparison exercise and identify which one is efficient. The competing models include the standard models of GARCH (1,1) and SV with an AR(1) log-volatility process, as well as flexible models with jumps, volatility in mean, leverage effects, $t$ distributed and moving average innovations (Chan and Grant, 2016).

Third, rather than focusing on specific market periods, this paper employs a parametric quantile causality test that is powerful enough under all market conditions jointly (e.g., bearish and bullish market, low volatility and high volatility market). Therefore, we can examine the predictive ability of the U.S. partisan conflict index for the oil market and gold market under different market conditions. By doing so, it also allows us to see under what conditions the U.S. partisan conflict index predicts oil and gold prices.

Fourth, the parametric test of Granger causality in quantiles used in this paper allows us to investigate causal relationships at any chosen conditional quantiles without pre-selecting some arbitrary sub-samples (Jeong et al., 2012). Existing studies have shown that oil and gold prices have nonlinear and structural mutation characteristics (Chen et al., 2014; Kirkulak Uludag and Lkhamazhapov, 2014; Gil-Alana et al., 2015), which may have an adverse impact on linear model estimation (Troster et al., 2019). Traditionally, most of the literature chooses a segment the sample (Fan and Xu, 2011), but doing this may result in the loss of sample information. The approach we used has two novel aspects. (i) The approach takes different locations and scales of the conditional distribution into account, which can provide more rich information on causality between US partisan conflicts and commodities prices than the traditional mean causality. (ii) It can address the problem of structural breaks and sample segmentation.

The remainder of the paper is organized as follows. Section 2 provides a brief review of the relevant literature. Section 3 presents the data and preliminary analysis. Section 4 introduces the empirical methods and research framework. Section 5 discusses the empirical results. Section 6 summarizes the major conclusions of this study and provides some implications for policy.

## 2. Literature

Together with use as a diversifier, hedge and safe haven for different traditional investments, the sustainable source of financialization in commodity markets over the last decade have generated a great deal of interest in the predictability of commodity returns and volatilities (Zhou et al., 2019). As two important commodities in the world, the understanding of drivers of oil and gold prices has become a focus of attention and research in recent years. With deeper financialization of the crude oil and gold market, a larger body of literature focuses on the non-fundamental factors, such as the EPU, investor attention, and political risks. In this section, the impacts of non-fundamental factors on the prices of crude oil and gold are discussed.

For the oil market, Hamilton (2009) analyzes the cause and influence of oil price shocks between 2007 and 2008 and concludes that most of the shocks are triggered by politics-induced oil production halts. Aloui et al. (2016) use a copula approach and find that higher uncertainty, measured by equity and EPU indices, significantly increases crude oil returns only during specific periods. Balcilar et al. (2017a) find U.S. EPU and equity market uncertainty has strong predictive power for oil returns over the entire distribution barring regions around the median; for oil volatility, the predictability virtually covers the entire distribution. By using an SVAR model, Chen et al. (2016) find support that political risk from OPEC countries has a significant and positive influence on Brent oil prices. Uddin et al. (2018) use the EPU index of US and Europe, implied US bond volatility index, the US VIX index, market sentiment and US speculation index as geopolitical risk, and estimate the impact of geopolitical risks on the oil price changes. Qadan and Nama (2018) show that investor sentiment, captured by nine different proxies such as the U.S. EPU index and consumer confidence index, has a significant effect on the returns and volatility of oil prices. Investor attention in the oil market has been found to hve a strong predictive power on oil prices (Yao et al., 2017; Han et al., 2017; Afkhami et al., 2017).

With regard to the gold market, Jones and Sackley (2016) incorporate an index of the U.S. and European EPU into a gold-pricing model and find that gold prices are positively related to EPU. The most frequently used method in analyzing the gold market is a nonparametric causality-in-quantiles test (Balcilar et al., 2016; Baur and Dimpfl, 2016; Li and Lucey, 2017; Bilgin et al. ,2018; Raza et al., 2018). For example,

Balcilar et al. (2016) confirm that policy and equity-market uncertainty can affect gold-price returns and volatility. Baur and Dimpfl (2016) choose internet search queries for gold as the investor's attention to gold price movements and find a positive relationship between gold price volatility and search queries. Li and Lucey (2017) find that EPU is a positive determinant of gold prices being a safe investment haven. Balcilar et al. (2017b) show that the effect of investor sentiment is more prevalent on intraday volatility in the gold market, rather than daily returns, while Bilgin et al. (2018) and Raza et al. (2018) demonstrate the importance of EPU in estimating gold prices.

Some studies focus on the relationship between commodities prices and financial market uncertainties. For example, Bouri et al. (2017a) find that there exist cointegration relationships and a nonlinear and positive impact of the implied volatilities of gold and oil on the implied volatility of the Indian stock market. Interestingly, Ji et al. (2018) investigate the information flow among US equities, strategic commodities and BRICS equities, and find that political, war, macroeconomic and financial events impact the changes in information flow among implied volatility indices. Dutta et al. (2019) examine the cointegration and nonlinear causality among the crude oil market and precious metal markets (gold and silver) and gold-miner stocks and find a bidirectional and symmetric effect between crude oil and gold markets.

Other literature studies the hedging abilities of energy and non-energy commodities. Bouri et al. (2017b) find significant bi-directional effects between gold and the Chinese and Indian stock markets in both high and low frequencies, suggesting that the safe-haven property of gold is not stable. Using a NARDL model, Rehman et al. (2019) find that the presence of short- and long-run asymmetric relationships between energy, non-energy commodities, and crude oil, offers more diversification benefits when combined with gold or silver.

To sum up, a large body of literature focuses on the relationship between financial markets and the commodities market, the hedging abilities of energy and non-energy commodities and influential factors behind strategic commodity prices. In particular, explaining the influencing factors behind strategic commodities has become a hot topic in recent years. We find that most studies examine the determinants of oil and gold prices from non-fundamental factors such as geopolitical risk, investor sentiment or attention and EPU, yet little research has investigated the effect of U.S. partisan conflict on gold prices. To this end, this paper estimates of the strategic commodities price response to the U.S. partisan conflict by employing a parametric test of Granger causality in quantiles, and provides policy suggestions for investors and policy marketers.

## 3. Methodology

Our empirical approach is twofold. First, two classes of time-varying volatility models, namely the GARCH models and the SV models are used to estimate the conditional volatility of oil and gold. Then, we examine which model works better to model the price volatility of oil and gold by comparing the Bayes factor. Second, a novel parametric test of Granger causality in quantiles proposed by Troster (2018) is employed to investigate the ability of the U.S. partisan conflict index to predict the returns and volatility of oil and gold prices.

3.1. Time-varying volatility models: GARCH and SV models

The competing models include the standard models of GARCH (1,1) and SV with an AR (1) log-volatility process, as well as flexible models with jumps, volatility in mean, leverage effects, and $t$ distributed and moving average innovations (for detailed model descriptions see Chan and Grant (2016)). Both the GARCH and SV models are estimated using Bayesian techniques. Joint sampling of the log volatilities is a crucial step to estimate the SV models. Based on the precision sampler of Chan and Jeliazkov (2009), estimating the SV models is done through the acceptance-rejection Metropolis-Hastings algorithm described in Chan (2017).

Following Chan and Grant (2016), we introduce the Bayesian model comparison via the Bayes factor and outline an efficient approach to compute the Bayes factor using importance sampling (see Table 1). Since the Bayes factor is simply a ratio of two marginal likelihoods, researchers often only report the marginal likelihoods of the set of competing models (Chan and Grant, 2016). We list the marginal likelihoods of different volatility models in this paper. If the observed data are likely under the model $M_1$, the associated marginal likelihood would be "largest" among these competing models.

**Table 1** GARCH models and SV models

| Pane 1: GARCH models | |
|---|---|
| GARCH | GARCH(1,1) model where $\sigma_2^t$ follows a stationary AR(1) |
| GARCH-2 | Same as GARCH but follows $\sigma_2^t$ a stationary AR(2) |
| GARCH-J | Same as GARCH but the returns equation has a "jump" component |
| GARCH-M | Same as GARCH but $\sigma_2^t$ enters the returns equation as a covariate |
| GARCH-MA | Same as GARCH but the observation error follows an MA(1) |
| GARCH-t | Same as GARCH but the observation error follows a t distribution |
| GARCH-GJR | GARCH with a leverage effect |
| **Panel 2: SV models** | |
| SV | SV model where $h_t$ follows a stationary AR(1) |
| SV-2 | Same as SV but $h_t$ follows a stationary AR(2) |
| SV-J | Same as SV but the returns equation has a "jump" component |
| SV-M | Same as SV but $h_t$ enters the returns equation as a covariate |
| SV-MA | Same as SV but the observation error follows an MA(1) |
| SV-t | Same as SV but the observation error follows a t distribution |
| SV-L | SV with a leverage effect |

Notes: $\sigma_2^t$ and $h_t$ denote the conditional variance in GARCH and SV models, respectively. For detailed model descriptions, see Chan and Grant (2016).

### 3.2. Granger causality in quantiles

We use the Granger causality test in quantiles proposed by Troster (2018) and Troster et al. (2019) to examine the ability of the U.S. partisan conflict index in predicting the returns and volatility of oil and gold prices across different conditional quantiles. Suppose U.S. partisan conflict index is $Z_t$ and the returns and volatility of the oil and gold is $Y_t$, respectively.

According to Granger (1969), a series $Z_t$ does not Granger-cause another series $Y_t$ if the past $Z_t$ does not help to predict future $Y_t$ given past $Y_t$. Suppose explanatory vector $I_t \equiv (I_t^{Y'}, I_t^{Z'})' \in R^d$, $d = s+q$, where $I_t^Y := (Y_{t-1},...Y_{t-s})' \in R^s$ and $I_t^Z := (Z_{t-1},...Z_{t-q})' \in R^q$. The null hypothesis of Granger non-causality from $Z_t$ to $Y_t$ as follows:

$$H_0^{Z \mapsto Y}: F_Y\left(y|I_t^Y, I_t^Z\right) = F_Y\left(y|I_t^Y\right), \quad \text{for all } Y \in R. \tag{1}$$

where $F_Y(y|I_t^Y, I_t^Z)$ and $F_Y(y|I_t^Y)$ be the conditional distribution functions of $Y_t$ given $(I_t^Y, I_t^Z)$ and $I_t^Y$, respectively. Test Granger non-causality in mean, which is only a necessary condition for (1). In this case, $Z_t$ does not Granger cause $Y_t$ in mean if

$$E(Y_t|I_t^Y, I_t^Z) = E(Y_t|I_t^Y) \quad \text{a. s.}, \tag{2}$$

where $E(Y_t|I_t^Y, I_t^Z)$ and $E(Y_t|I_t^Y)$ are the mean of $F_Y(\cdot|I_t^Y, I_t^Z)$ and $F_Y(\cdot|I_t^Y)$

respectively. Granger non-causality in mean can be easily extended to higher-order moments. However, causality in mean overlooks the dependence that may appear in conditional tails of the distribution. Thus, a test Granger non-causality in conditional quantiles is proposed. Let $Q_\tau^{Y,Z}\left(\cdot|I_t^Y,I_t^Z\right)$ be the τ-quantiles of $F_Y\left(\cdot|I_t^Y,I_t^Z\right)$, we rewrite Eq. (1) as follows:

$$H_0^{QC:Z\mapsto Y}: Q_\tau^{Y,Z}\left(Y_t|I_t^Y,I_t^Z\right)=Q_\tau^Y\left(Y_t|I_t^Y\right), \text{ a. s. for all } \tau\in T, \tag{3}$$

where $\Gamma$ is a compact set such that $\Gamma\subset[0,1]$ and the conditional $\tau$-quantiles of $Y_t$ satisfying the following restrictions:

$$\Pr\left\{Y_t\leq Q_\tau^Y\left(Y_t|I_t^Y\right)|I_t^Y\right\}:=\tau, \text{ a. s. for all } \tau\in T,$$

$$\Pr\left\{Y_t\leq Q_\tau^{Y,Z}\left(Y_t|I_t^Y,I_t^Z\right)|I_t^Y,I_t^Z\right\}:=\tau, \text{ a. s. for all } \tau\in T \tag{4}$$

Given an explanatory vector $I_t$, we have $\Pr\left\{Y_t\leq Q_\tau\left(Y_t|I_t\right)|I_t\right\}=E\left\{\mathrm{I}\left[Y_t\leq Q_\tau\left(Y_t|I_t\right)\right]|I_t\right\}$, where $I(Y_t\leq y)$ is an indicator function of the event that a is less or equal than $y$. Thus (4) is equivalent to

$$\left\{\mathrm{I}\left[Y_t\leq Q_\tau^{Y,Z}\left(Y_t|I_t^Y,I_t^Z\right)\right]|I_t^Y,I_t^Z\right\}=E\left\{\mathrm{I}\left[Y_t\leq Q_\tau^Y\left(Y_t|I_t^Y\right)\right]|I_t^Y,I_t^Z\right\},$$

$$\text{a. s. for all } \tau\in T, \tag{5}$$

where the left-hand side of (5) is equal to the $\tau$-quantile of $F_Y\left(\cdot|I_t^Y,I_t^Z\right)$ by definition. Following Troster (2018), we postulate a parametric model to estimate the $\tau th$ quantile of $F_Y\left(\cdot|I_t^Y\right)$, where we assume that $Q_\tau\left(\cdot|I_t\right)$ is correctly specified by a parametric model $m(\cdot,\theta(\tau))$ belonging to a family of functions $\mathrm{M}=\left\{m(\cdot,\theta(\tau))|\theta(\cdot)\tau\mapsto\in\Theta\subset R^P, \tau\in T\subset[0,1]\right\}$. Let $\mathrm{B}\subset\mathrm{M}$ be a family of uniformly bounded functions $\tau\mapsto\theta(\tau)$ such that $\theta(\tau)\in\Theta\subset R^P$. Then, under the null hypothesis in (3), the τ-conditional quantile $Q_\tau^Y\left(\cdot|I_t^Y\right)$ is correctly specified by a parametric model $m\left(I_t^Y,\theta_0(\tau)\right)$, for some $\theta_0\in\mathrm{B}$, using only the restricted information set $I_t^Y$, and we redefine our testing problem in (3) as

$$H_0^{Z\mapsto Y}: E\left[\mathrm{I}\left(Y_t\leq m\left(I_t^Y,\theta_0(\tau)\right)\right)|I_t^Y,I_t^Z\right]=\tau, \text{ a. s. for all } \tau\in T, \tag{6}$$

versus:

$$H_A^{Z \mapsto Y}: E\left[I\left(Y_t \leq m\left(I_t^Y, \theta_0(\tau)\right)\right) \middle| I_t^Y, I_t^Z\right] \neq \tau, \text{ for some } \tau \in T, \tag{7}$$

where $m(I_t^Y, \theta_0(\tau))$ correctly specifies the true conditional quantile $Q_\tau^Y(\cdot | I_t^Y)$, for all $\tau \in T$. We rewrite (6) as $H_0^{Z \mapsto Y}: E\left\{\left[I\left(Y_t - m\left(I_t^Y, \theta_0(\tau)\right) \leq 0\right) - \tau\right] \middle| I_t^Y, I_t^Z\right\} = 0$ almost surely, for all $\tau \in \Gamma$. Then we can characterize the null hypothesis (6) by a sequence of unconditional moment restrictions:

$$H_0^{Z \mapsto Y}: E\left\{\left[I\left(Y_t - m\left(I_t^Y, \theta_0(\tau)\right) \leq 0\right) - \tau\right] \exp(i\varpi' I_t)\right\} = 0 \tag{8}$$

Where $\exp(i\varpi' I_t) = \exp\left[i\left(\varpi_1(Y_{t-1}, Z_{t-1})' + \ldots + \varpi_r(Y_{t-r}, Z_{t-r})'\right)\right]$ is a weighting function, for all $\varpi \in R^d$ with $r \leq d$, and $i = \sqrt{-1}$ is the imaginary root. The test statistic is a sampled analogue of $E\left\{\left[I\left(Y_t - m\left(I_t^Y, \theta_0(\tau)\right) \leq 0\right) - \tau\right] \exp(i\varpi' I_t)\right\}$:

$$v_T(\varpi, \tau) := \frac{1}{\sqrt{T}} \sum_{t=1}^{T} \left[I\left(Y_t - m\left(I_t^Y, \theta_0(\tau)\right) \leq 0\right) - \tau\right] \exp(i\varpi' I_t) \tag{9}$$

where $\theta_T$ is a $\sqrt{T}$-consistent estimator of $\theta_0(\tau)$, for all $\tau \in T$. Then, we apply the test statistic:

$$S_T := \int_T \int_W |v_T(\varpi, \tau)|^2 dF_\varpi(\varpi) dF_\tau(\tau) \tag{10}$$

where $F_\varpi(\cdot)$ is the conditional distribution function of the ad-variate standard normal random vector, $F_\tau(\cdot)$ as a uniform discrete distribution over a grid of T in $n$ equidistributed points, $T_n = \{\tau_j\}_{j=1}^{n}$. And the vector of weights $\varpi \in R^d$ is drawn from a standard normal distribution. The test statistic in (10) can be estimated using its sample counterpart. Let $\Psi$ be the $T \times n$ matrix $\Psi$ with elements $\psi_{i,j} = \Psi_{\tau_j}(Y_i - m(I_i^Y, \theta_T(\tau_j)))$. Then, the test statistic $S_T$ has the form

$$S_T = \frac{1}{Tn} \sum_{j=1}^{n} |\psi'_{\cdot j} \mathbf{W} \psi_{\cdot j}| \tag{11}$$

where $\mathbf{W}$ is the $T \times T$ matrix with elements $w_{t,s} = \exp\left[-0.5(I_t - I_s)^2\right]$, and $\psi_{\bullet j}$ denotes the *jth* column of $\Psi$. It rejects the null hypothesis whenever it observes "large" values of $S_T$.

We use the subsampling procedure of Troster (2018) to calculate critical values for $S_T$ in Eq. (11). Given our series $\{X_t = (Y_t, Z_t)\}$ of sample size $T$, we generate $B = T - b + 1$ subsamples of size $b$ (taken without replacement from the original data) of the form $\{X_i, \ldots, X_{i+b-1}\}$. Then, the test statistic $S_T$ in Eq. (11) is calculated for each subsample; we obtain p-values by averaging the subsample test statistics over the $B$ subsamples. Following Troster (2018), we choose a subsample of size $b = [kT^{2/5}]$, where $[\cdot]$ is the integer part of a number, and k is a constant parameter. To apply the $S_T$ test in Eq. (11), we specify three different quantile auto regressive (QAR) models $m(\cdot)$, for all $\tau \in \Gamma \subset [0,1]$, under the null hypothesis of non-Granger-causality in Eq. (9) as follows:

$$QAR(1): m^1(I_t^Y, \theta(\tau)) = \mu_1(\tau) + \mu_2(\tau)Y_{t-1} + \sigma_t \Phi_u^{-1}(\tau)$$

$$QAR(2): m^2(I_t^Y, \theta(\tau)) = \mu_1(\tau) + \mu_2(\tau)Y_{t-1} + \mu_3(\tau)Y_{t-2} + \sigma_t \Phi_u^{-1}(\tau)$$

$$QAR(3): m^3(I_t^Y, \theta(\tau)) = \mu_1(\tau) + \mu_2(\tau)Y_{t-1} + \mu_3(\tau)Y_{t-2} + \mu_4(\tau)Y_{t-3} + \sigma_t \Phi_u^{-1}(\tau) \quad (12)$$

where the parameters $\theta(\tau) = (\mu_1(\tau), \mu_2(\tau), \mu_3(\tau), \mu_4(\tau), \sigma_t)'$ are estimated by maximum likelihood in an equally spaced grid of quantiles and $\Phi_u^{-1}(\cdot)$ are the inverse of a standard normal distribution function. To verify the signature of the causal relationship between the variables, we estimate the quantile autoregressive models in Eq. (12) including lagged variables of another variable. For simplicity, we present the results using only a QAR (3) model with the lagged values of the other variable as follows:

$$Q_\tau^{Y,Z}\left(Y_t \mid I_t^Y, I_t^Z\right) = \mu_1(\tau) + \mu_2(\tau)Y_{t-1} + \mu_3(\tau)Y_{t-2} + \mu_4(\tau)Y_{t-3} + \beta(\tau)Z_{t-1} + \sigma_t \Phi_u^{-1}(\tau) \quad (13)$$

## 4. Data

This paper uses monthly data from January 1981 to October 2017 to investigate the ability of the U.S. partisan conflict index to predict the returns and volatility of oil and gold prices. The U.S. partisan conflict index, developed by Azzimonti (2018), is obtained from the website of the Federal Reserve Bank of Philadelphia[3]. This index tracks the degree of political disagreement among U.S. politicians at the federal level by measuring the frequency of newspaper articles reporting disagreement in a given month. Higher index values indicate greater conflict among the political parties, congress, and the president. The US refiner's acquisition cost of crude oil is used as the measure of the global crude oil prices, was obtained from the US Energy Information Administration (EIA) in the United States Department of Energy[4]. The gold price is the Gold Fixing Price at 3:00 P.M. (London time) in the London Bullion Market based in U.S. Dollars, and was obtained from the FRED database of the Federal Reserve Bank of St. Louis[5]. Oil and gold prices are divided by the US consumer price index (CPI) from the Bureau of Labor Statistics to obtain the real price. Returns for oil and gold are measured by the first-differenced natural logarithm of the real oil and gold prices.

---

[3] https://www.philadelphiafed.org/research-and-data/real-time-center/partisan-conflict-index

Summary statistics of the variables are reported in Table 2. Note that oil returns have higher volatility than gold returns. Oil returns are skewed to the left, and gold returns and U.S. partisan conflict index skewed to the right, with all the variables having excess kurtosis. The Jarque–Bera test overwhelmingly rejects the null of normality. This evidence of fat tails in the variables provides us with the preliminary motivation to use a causality-in-quantile test rather than standard linear Granger causality test based on the conditional mean.

**Table 2** Summary statistics

|  | *Gold* | *Oil* | *partisan* |
| --- | --- | --- | --- |
| Mean | -0.064 | -0.119 | 4.627 |
| Median | -0.322 | 0.196 | 4.572 |
| Maximum | 17.587 | 36.641 | 5.603 |
| Minimum | -15.194 | -33.942 | 4.081 |
| Std. Dev. | 3.895 | 7.437 | 0.266 |
| Skewness | 0.162 | -0.552 | 0.811 |
| Kurtosis | 5.180 | 6.743 | 3.580 |
| Jarque-Bera | 89.468*** | 280.452*** | 54.598*** |
| Observations | 442 | 442 | 442 |

Notes: Std.Dev denotes standard deviation; *Gold* and *Oil* represent the gold returns and crude oil returns, respectively. *partisan* denotes the U.S partisan conflict index taken in natural logarithm. *** denotes the rejection of the null of normality of the Jarque-Bera test at 1% level of significance.

Concerning the volatility evaluation for oil and gold market, seven GARCH models and seven SV models are estimated by using Bayesian techniques (see Chan and Grant (2016)). The marginal likelihoods are computed using the improved cross-entropy method of Chan and Eisenstat (2015). Model comparison results are reported in Table 3. Overall, results suggest that the best model for modelling the price volatility of oil and gold is the SV-MA model. The marginal likelihood of the SV-MA model for estimating the volatility of crude oil and gold prices is the largest, -1393.9 and -1196.2, respectively. Therefore, in this paper, we employ the SV-MA model to measure the price volatility of crude oil and gold. This finding of oil volatility estimation is in line with Chan and Grant (2016) who found that the SV-MA model is best for estimating oil price volatility.

---

[4] www.eia.gov/.
[5] https://research.stlouisfed.org/fred2/

Furthermore, we investigate which features are essential in modelling the dynamic volatility of oil and gold prices by comparing the different GARCH and SV models. By comparing the GARCH with the GARCH-2 and the SV with the SV-2, we conclude that the model with AR(2) volatility process provides a higher marginal likelihood. Thus, the volatility models with the AR(2) volatility process are better than the models with AR (1) volatility process.

Next, we examine the importance of volatility feedback for modelling price volatility of oil and gold. We find that adding the volatility feedback component markedly increases the marginal likelihood of the SV model for oil and gold. By comparison, the model-fit cannot be improved by adding the volatility feedback component in the GARCH for the gold prices.

To investigate the relevance of the moving average component, we compare the GARCH with the GARCH-MA and the SV with the SV-MA. For SV models, adding the MA component improves the model-fit for crude oil and gold prices. However, for GARCH models, the marginal likelihood cannot be improved by adding the MA component for crude oil and gold.

We also examine whether the GARCH model and SV model with the "jump" can better shape the price volatility of crude oil and gold than the conventional GARCH and SV model. It shows that the marginal likelihood has improved when adding the jump component into GARCH and SV models. Therefore, it is essential to consider the jump component in the GARCH model and SV model when estimating the price volatility in the crude oil and gold market.

Finally, by comparing the GARCH with the GARCH-GJR and the SV with the SV-L, we conclude that the leverage effect is vital for modelling price volatility of crude oil and gold compared to the conventional GARCH and SV model without the specification of leverage. As we know, the leverage effects are essential for stock returns and crude oil prices (Chan and Grant, 2016). These results support the argument of the 'financialization' of the crude oil and gold market.

**Table 3** Log marginal likelihoods of GARCH and SV models for the crude oil and gold.

|  | *Oil* | *Gold* |
|---|---|---|
| GARCH | -1460.3 | -1213.3 |
|  | (0.02) | （0.02） |
| SV | -1433.6 | -1200.2 |
|  | (0.03) | （0.02） |
| GARCH-2 | -1458.7 | -1213.6 |
|  | (0.04) | -0.06 |
| SV-2 | -1431.2 | -1201.3 |
|  | (0.08) | (0.03) |
| GARCH-J | -1434.9 | -1205.7 |
|  | (0.20) | (0.08) |
| SV-J | -1434.5 | -1200.1 |
|  | (0.05) | (0.09) |
| GARCH-M | -1446.8 | -1220.0 |
|  | (0.04) | (0.03) |
| SV-M | -1440.6 | -1205.4 |
|  | (0.03) | (0.02) |
| GARCH-MA | -1427.1 | -1204.5 |
|  | (0.04) | (0.03) |
| **SV-MA** | **-1393.9** | **-1196.2** |
|  | **(0.03)** | **(0.01)** |
| GARCH-t | -1434.1 | -1197.3 |
|  | (0.01) | (0.01) |
| SV-t | -1434.0 | -1195.7 |
|  | (0.02) | (0.02) |
| GARCH-GJR | -1457.4 | -1215.7 |
|  | (0.02) | (0.03) |
| SV-L | -1455.9 | -1201.6 |
|  | (0.02) | (0.02) |

Notes: Table 3 displays the values of log marginal likelihoods for oil and gold volatility estimation using GARCH and SV models. The absolute value of this indicator is smaller, indicating the model is better. The numerical standard errors are in parentheses.

We perform standard unit root tests to determine whether the returns and volatility of oil and gold and U.S. partisan conflict index series are stationary. Test results are reported in Table 4. According to results in Table 4, the Augmented-Dickey and Fuller (ADF) test (Dickey and Fuller, 1979) and the Phillips-Perron (PP) test (Phillips and Perron, 1988) reject the null hypothesis of non-stationarity for all series. However, a major shortcoming with the standard unit root tests is that they do not allow for the possibility of structural breaks. Therefore, we use the unit root test proposed by Perron (1997), which allows a break at an unknown location on both the trend and the intercept for any variables. We evaluate the unit root test for the returns and volatility of oil and gold, and U.S. partisan conflict index. The results of the Perron (1997) unit root test and the estimated break date are also shown in Table 4. The Perron unit root tests confirm these series are stationary, and there exists a break for oil returns, gold returns and partisan conflict index in December 2008, September 1982 and December 2009, respectively. Meanwhile, it detects that the oil volatility and gold volatility have a break in October 2008 and September 1982, respectively. This finding of breakpoints in the returns and volatility of oil and gold and partisan conflict index indicates that the linear model based on mean estimation are not suitable to depict the relationship between them.

**Table 4** Unit root test

|  | ADF | | PP | | Perron test with break | |
| --- | --- | --- | --- | --- | --- | --- |
|  | C | C+T | C | C+T | C+T | date |
| *Conflict* | -2.623*(3) | -3.096***(3) | -5.357***(6) | -6.637***(8) | 6.229***(3) | 2009m12 |
| *Oil* | -12.045***(1) | -12.054(1) | -10.869(17) | -10.803***(18) | -12.562***(2) | 2008m12 |
| *Gold* | -18.243***(0) | -18.446***(0) | -18.265***(8) | -18.402***(7) | -19.261***(0) | 1982m9 |
| *Oil-VOL* | -4.006***(1) | -4.487***(1) | -2.942**(12) | -3.311(12) | -5.864***(1) | 2008m10 |
| *Gold-VOL* | -3.242**(1) | -3.152*(1) | -2.593*(14) | -2.475(14) | -4.900**(1) | 1982m9 |

Notes: C denotes constant, T denotes trend; **and ***indicate significance at the 5% and 1% level, respectively. *Oil* and *Gold* represent the returns of oil and gold, respectively. *Oil-VOL* and *Gold-VOL* denote the price volatility of oil and gold based on the estimation of SV-MA model, respectively. The numbers in parentheses are the optimal lag order in the ADF and PP test based on the Schwarz Info criterion and Newey-west bandwidth.

## 5. Empirical results and discussions

### 5.1. Linear Granger causality test

Though our objective is to analyze the quantile causality running from the U.S. partisan conflict index to the returns and volatility of oil and gold prices, for the sake of completeness and comparability, we also conducted the standard linear Granger causality test (Granger, 1969) based on the VAR model. The lag parameters for the VAR model are selected based on the Akaike information criterion (AIC). Table 5 presents the results for the linear Granger causality test. The null hypothesis of non-causality from the U.S. partisan conflict index to the returns of oil and gold cannot be rejected at the 10% significance level. For the price volatility of gold, the null hypothesis of non-causality from the U.S. partisan conflict index to the volatility of gold can be rejected at the 5% significance level. However, for oil price volatility, the null hypothesis cannot be rejected. These results may be due to the misspecification of the test model. It is well-known that the linear Granger causality test can miss the important nonlinear causal relationship (Jiang et al., 2019). Therefore, the insufficient or weak evidence for the causal relationship can be attributed to the low power of the linear Granger causality test if the time series analyzed is nonlinear or non-normal.

**Table 5** Linear Granger causality test (causality between the U.S. partisan conflict and the returns and volatility for oil and gold market)

| Null hypothesis | Lag | Chi-sq | P-value | Causality or not |
|---|---|---|---|---|
| **Panel 1:** For oil and gold returns | | | | |
| $partisam \mapsto oil$ | 8 | 9.003 | 0.342 | NO |
| $oil \mapsto partisam$ | 8 | 14.567* | 0.068 | Yes |
| $partisam \mapsto gold$ | 8 | 7.686 | 0.465 | NO |
| $gold \mapsto partisam$ | 8 | 10.738 | 0.217 | NO |
| **Panel 2:** For oil and gold price volatility | | | | |
| $partisam \mapsto oil$ | 8 | 9.052 | 0.338 | No |
| $oil \mapsto partisam$ | 8 | 3.544 | 0.896 | NO |
| $partisam \mapsto gold$ | 8 | 16.886** | 0.031 | Yes |
| $gold \mapsto partisam$ | 8 | 11.019 | 0.201 | No |

Notes: *, **and ***indicate significance at the 10%, 5% and 1% level, respectively; the symbol $\mapsto$ represents the null hypothesis of no linear causality. The lag parameters are selected based on the AIC. Yes in the last column indicates that the null hypothesis was rejected at least at the 10% significance level.

### 5.2. Non-linear Granger causality tests

Given the strong evidence of nonlinearity obtained from the BDS tests (see Table A.1 in Appendix A), we further investigate whether there exists nonlinear Granger causality running from the U.S. partisan conflict index to the returns and volatility of oil and gold prices. To this end, we use the nonlinear Granger causality test of the H&J test (Hiemstra and Jones, 1994) and the D&P test (Diks and Panchenko, 2006). The results for the H&J test and D& P nonlinear Granger causality test are presented in Table 6. We perform the tests for the embedding dimension $m = 1.5$ and select the lags 1-6. It is noticeable that the null hypothesis of no nonlinear Granger causality running from the U.S. partisan conflict index to the returns and volatility of oil and gold prices in the sample period cannot be rejected at the 10% significance level. For these findings of no causality, it is because the nonlinear Granger causality test approaches rely on conditional-mean based estimation, and fail to capture the entire conditional distribution of returns and volatility of oil and gold prices. Given the nonexistence of any evidence of the nonlinear Granger causality, we next turn to causality-in-quantiles tests, which considers all quantiles of the distribution not only the center of the distribution. It can provide more detail information on the relationship between the U.S. partisan conflict and oil and gold price movements for investors.

**Table 6** Non-linear Granger causality tests (causality between the U.S. partisan conflict and the returns and volatility for oil and gold market)

| $l_X = l_Y$ | H&J (P-value) | D&P (P-value) | H&J (P-value) | D&P (P-value) |
|---|---|---|---|---|
| Panel 1 *Returns* | partisam $\mapsto$ oil returns | | partisam $\mapsto$ gold returns | |
| 1 | 0.732 (0.231) | 0.714(0.237) | 0.687 (0.245) | 0.665 (0.252) |
| 2 | 0.3179 (0.375) | 0.233(0.407) | 0.443 (0.328) | 0.329 (0.371) |
| 3 | 0.468 (0.319) | 0.405(0.342) | 0.693 (0.243) | 0.601 (0.273) |
| 4 | 1.388*(0.082) | 1.149(0.125) | 0.763 (0.222) | 0.768 (0.221) |
| 5 | 1.531*(0.062) | 1.321*(0.093) | 0.684 (0.246) | 0.641 (0.261) |
| 6 | 1.361*(0.086) | 1.145(0.126) | 0.345 (0.364) | 0.393 (0.346) |
| Panel 2: *Volatility* | partisam $\mapsto$ oil volatility | | partisam $\mapsto$ gold volatility | |
| 1 | -1.090 (0.862) | -0.916 (0.820) | -0.629 (0.735) | -0.501 (0.691) |
| 2 | -1.570 (0.941) | -1.248 (0.894) | -0.544 (0.706) | -0.246 (0.597) |
| 3 | -1.774 (0.961) | -1.343 (0.910) | 0.761 (0.776) | -0.315 (0.623) |
| 4 | -1.931 (0.973) | -1.411 (0.920) | -0.684 (0.753) | -0.164 (0.565) |
| 5 | -1.498 (0.932) | -0.957 (0.830) | -0.727 (0.766) | -0.091 (0.536) |
| 6 | -1.198 (0.884) | -0.702 (0.758) | -0.802 (0.788) | -0.083 (0.532) |

Notes: *, **and ***indicate rejection of the null hypothesis at the 10%, 5% and 1% level, respectively; $l_X = l_Y$ denotes the lag length. Symbol $\mapsto$ represents no nonlinear causality.

### 5.3. Granger causality test in quantiles

In this section, we analyze the importance of U.S. partisan conflict in predicting the returns and volatility of oil and gold market over the entire conditional distribution of oil returns and volatility by employing a causality test in quantiles proposed by Troster (2018) (Results in Tables 7 and 8). Troster (2018) built a test statistic $S_T$ and proposed a subsampling procedure to calculate critical values $S_T$. To apply the $S_T$ test, three different QAR models are estimated for each dependent variable at lag length from one to three, respectively.

### 5.3.1. Causality from U.S. partisan conflict to returns

Table 7 presents the p-values for the test of quantile-causality running from the U.S. partisan conflict index to the returns of strategic commodities (crude oil and gold). Overall, it shows that the quantile-causality test for the quantile interval [0.1, 0.9] is significant at the 10% significance level, indicating that the U.S. partisan conflict has a strong ability to predict the returns of oil and gold prices. Moreover, it is mainly found that there is an outstanding pattern of Granger-causality from the U.S. partisan conflict index to the returns of oil and gold prices in the tails of the distribution of the strategic commodities price movements.

More specifically, as for oil returns, the test results of causality-in-quantiles running from partisan conflict to oil returns are insignificant at the median (quantiles at 0.5) but become significant at the tail quantiles of the conditional distribution of oil returns. The insignificance of the test results at the median of the conditional distribution of oil returns is in line with the results of the conditional mean estimation analysis of Table 5 and Table 6 (the linear Granger causality test and nonlinear Granger causality test, respectively) which does not find any evidence that (relative) partisan conflict predicts oil returns as well. This finding is partly in line with Balcilar et al. (2017a) who find that U.S. EPU and equity market uncertainty have strong predictive power for oil returns over the entire distribution barring regions around the median. Likewise, Aloui et al. (2016) and Shahzad et al. (2017) confirm that there exist significant causal-flows from the U.S. EPU to oil returns over the entire sample period and for the majority of the quantile ranges as well. Furthermore, different from their studies, we find that the effect of U.S. partisan conflict on the conditional distribution of oil returns is particularly significant for the lower quantiles, viz. at around 0.1, 0.2, 0.3 and 0.4 quantiles. However, for oil returns at the higher quantiles such as at quantiles 0.6, 0.7, and 0.8, the partisan conflict does affect oil returns, with the 0.9 quantiles being an exception. This implies that the explanatory power of partisan conflict on oil returns is heterogeneous in different market conditions. Specifically, when the crude oil market is in a bearish state (oil returns at the lower quantiles), the partisan conflict has a significant impact on the oil returns. However, under a bullish state (oil returns at the higher quantiles), the impact of the partisan conflict on oil returns is limited. The possible reason behind the finding is that partisan conflict in the U.S. increases the uncertainty of future policy, thereby resulting in a decrease in oil demand. When crude oil is in a bear market, that is, the price of crude oil is very low, the reduction in oil demand caused by U.S. partisan conflict will more likely lead to a decrease in oil prices than that in a bullish market.

Turning to gold returns, as can be seen from Table 7, the rejection of the null hypothesis of no causality running from partisan conflict to gold returns is concentrated more around the tail quantiles. Likewise, there is no evidence in favour of the predictable effect of partisan conflict on the gold returns at the median quantiles. The results confirm the finding of Balcilar (2016) and Jones and Sackley (2016) that there exists a causality running from U.S. EPU to gold returns. Li and Lucey (2017) find that EPU is a positive determinant of gold being a safe investment haven. This finding also supports the conclusion of Bilgin et al. (2018) that when using U.S. partisan conflict index as the EPU, worsening EPU contributes to increases in the price of gold. Not only that, we find that, for gold returns at lower quantiles, we only find the partisan conflict can affect the gold returns at 0.2 quantiles of the conditional distribution of gold returns, and for other lower quantiles, it is not valid.

In comparison, for gold returns at higher quantiles, the null hypothesis of no causality can be rejected around 0.6, 0.7, 0.8 and 0.9 quantiles. It indicates that the U.S. partisan conflicts are more likely to affect the higher quantiles of the conditional distribution of gold returns. In other words, partisan conflict matters only when the gold market is performing above its normal (average) mode, i.e., in bullish scenarios. This finding is different from the conclusions in the oil market. Media reports and investment recommendations often emphasize that gold acts as a classic safe-haven and hedging investment in times of economic and political uncertainty. Therefore, it is not surprising that when U.S partisan conflict intensifies, investors will choose gold to avoid risk, which has an impact on the gold market. Especially when the gold market is in a bull market, investors are more likely to choose gold to hedge their risk than that in a bear market, because they can gain higher profits in the bull market.

**Table 7** Quantile causality test results (sub-sampling p-values) from U.S. partisan conflict to returns of oil and gold market.

|  | Lag | [0.1, 0.9] | 0.1 | 0.2 | 0.3 | 0.4 | 0.5 | 0.6 | 0.7 | 0.8 | 0.9 |
|---|---|---|---|---|---|---|---|---|---|---|---|
| *Oil* | 1 | **0.094** | **0.018** | **0.047** | **0.078** | **0.091** | 0.852 | 0.122 | 0.122 | 0.156 | **0.096** |
| *returns* | 2 | **0.091** | **0.016** | **0.016** | **0.068** | **0.081** | 0.763 | 0.102 | 0.102 | 0.174 | **0.096** |

|   | | | | | | | | | | |
|---|---|---|---|---|---|---|---|---|---|---|
|  | 3 | **0.096** | **0.042** | **0.057** | **0.073** | **0.089** | 0.938 | 0.107 | 0.130 | 0.419 | **0.089** |
| *Gold returns* | 1 | **0.003** | 0.167 | **0.013** | 0.154 | 0.685 | 0.286 | **0.003** | **0.036** | **0.003** | **0.049** |
|  | 2 | **0.003** | **0.047** | **0.008** | 0.130 | 0.547 | 0.266 | **0.003** | **0.008** | **0.003** | **0.008** |
|  | 3 | **0.005** | 0.234 | **0.003** | 0.284 | 0.508 | 0.396 | **0.003** | **0.003** | **0.003** | **0.016** |

Notes: Bold *p*-values denote rejection of the null hypothesis at least 10% significance level.

### 5.3.2. Causality from U.S. partisan conflict to volatility

The volatility of commodity prices is often regarded as an indicator for the calculation of hedging. Therefore, it is meaningful to study the determinants of commodity price volatility. In this section, we examine whether the U.S. partisan conflict index has predictive power for the volatility of oil and gold prices. Table 8 displays the results of causality in quantile test. We find that the U.S. partisan conflict index affects the volatility of oil and gold prices over the entire conditional distribution, i.e., at various phases of the oil and gold market. This empirical evidence is consistent with Balcilar et al. (2017a), who find that for oil volatility, the predictability of U.S. EPU and equity market uncertainty virtually covers the entire distribution.

**Table 8** Quantile causality test results (subsampling *p*-values) from U.S. partisan conflict to price volatility of oil and gold.

|  | Lag | [0.1, 0.9] | 0.1 | 0.2 | 0.3 | 0.4 | 0.5 | 0.6 | 0.7 | 0.8 | 0.9 |
|---|---|---|---|---|---|---|---|---|---|---|---|
| *Oil volatility* | 1 | **0.049** | **0.003** | **0.003** | 0.104 | 0.193 | 0.198 | 0.109 | **0.049** | **0.029** | 0.138 |
|  | 2 | **0.016** | **0.003** | **0.003** | **0.005** | **0.003** | **0.023** | **0.081** | **0.018** | **0.005** | **0.003** |
|  | 3 | **0.008** | **0.003** | **0.003** | **0.008** | **0.003** | **0.003** | **0.052** | **0.018** | **0.008** | **0.003** |
| *Gold Volatility* | 1 | **0.003** | **0.005** | **0.003** | **0.008** | **0.010** | **0.068** | 0.102 | **0.003** | **0.003** | **0.010** |
|  | 2 | **0.003** | **0.003** | **0.003** | **0.003** | **0.003** | **0.003** | 0.161 | **0.003** | **0.003** | **0.003** |
|  | 3 | **0.003** | **0.003** | **0.003** | **0.003** | **0.003** | **0.003** | **0.091** | **0.003** | **0.003** | **0.003** |

**Notes:** Bold *p*-values denote rejection of the null hypothesis at least 10% significance level

### 5.4. Robustness analysis

5.4.1 More quantiles to depict the market states

In this section, we discuss the robustness of our results. One important conclusion of this paper is that the effect of U.S. partisan conflict on the returns and volatility of oil and gold is clustered around the tail of the conditional distribution of returns. To this end, in the estimation process of the quantile- causality test, we set more numbers of quantiles with 0.05 step length. For lower quantiles, they are denoted by quantiles 0.05, 0.1, 0.15, 0.2, 0.25, 0.3, 0.35, 0.4; median quantiles including quantiles at 0.45, 0.5, higher quantiles contain the quantiles at 0.55, 0.6, 0.65, 0.7, 0.75, 0.8, 0.85, 0.9, 0.95. Re-estimating the Granger quantiles causality test, results can be seen in Tables 9 and 10, the empirical results suggest that our main findings are not changed with setting more quantiles.

**Table 9** Quantile causality test results (sub-sampling *p*-values) from U.S. partisan conflict to oil and gold returns in more quantiles.

|  | *Oil returns* | | | *Gold returns* | | |
| --- | --- | --- | --- | --- | --- | --- |
| Lag | 1 | 2 | 3 | 1 | 2 | 3 |
| [0.05, 0.95] | **0.094** | **0.091** | **0.096** | **0.004** | **0.003** | **0.005** |
| 0.05 | 0.128 | 0.474 | 0.383 | 0.635 | 0.297 | 0.299 |
| 0.1 | **0.018** | **0.016** | **0.042** | 0.167 | 0.147 | 0.234 |
| 0.15 | **0.039** | **0.026** | **0.044** | **0.060** | **0.026** | **0.008** |
| 0.2 | **0.047** | **0.016** | **0.057** | **0.013** | **0.008** | **0.003** |
| 0.25 | **0.078** | **0.055** | **0.065** | **0.031** | 0.339 | 0.112 |
| 0.3 | **0.078** | **0.068** | **0.073** | 0.154 | 0.130 | 0.284 |
| 0.35 | **0.083** | **0.070** | **0.078** | 0.685 | 0.638 | 0.797 |
| 0.4 | **0.091** | **0.081** | **0.089** | 0.685 | 0.547 | 0.508 |
| 0.45 | 0.128 | 0.276 | 0.398 | 0.401 | 0.292 | 0.273 |
| 0.5 | 0.852 | 0.763 | 0.938 | 0.286 | 0.266 | 0.396 |
| 0.55 | 0.135 | 0.078 | **0.089** | 0.148 | **0.021** | **0.047** |
| 0.6 | 0.122 | 0.102 | 0.107 | **0.003** | **0.003** | **0.003** |
| 0.65 | **0.096** | 0.112 | 0.117 | **0.003** | **0.010** | **0.018** |
| 0.7 | 0.122 | 0.102 | 0.130 | **0.036** | **0.008** | **0.004** |
| 0.75 | 0.125 | 0.133 | 0.138 | **0.003** | **0.003** | **0.003** |
| 0.8 | 0.156 | 0.174 | 0.419 | **0.003** | **0.003** | **0.003** |
| 0.85 | **0.089** | **0.091** | 0.091 | **0.003** | **0.003** | **0.003** |
| 0.9 | **0.096** | **0.096** | 0.089 | **0.049** | **0.008** | **0.016** |
| 0.95 | 0.167 | 0.289 | 0.102 | **0.049** | **0.044** | **0.057** |

Notes: see Table 7.

**Table 10** Quantile causality test results (sub-sampling *p*-values) from U.S. partisan conflict to price volatility of oil and gold market.

|  | *Oil volatility* | | | *Gold volatility* | | |
| --- | --- | --- | --- | --- | --- | --- |
| Lag | 1 | 2 | 3 | 1 | 2 | 3 |
| [0.05, 0.95] | **0.047** | **0.016** | **0.008** | **0.003** | **0.003** | **0.003** |
| 0.05 | **0.003** | **0.008** | **0.010** | **0.003** | **0.049** | **0.039** |
| 0.1 | **0.003** | **0.003** | **0.003** | **0.005** | **0.003** | **0.003** |
| 0.15 | **0.003** | **0.034** | **0.026** | **0.003** | **0.003** | **0.003** |
| 0.2 | **0.003** | **0.003** | **0.003** | **0.003** | **0.003** | **0.003** |
| 0.25 | **0.031** | **0.005** | **0.003** | **0.010** | **0.003** | **0.003** |
| 0.3 | 0.104 | **0.005** | **0.008** | **0.008** | **0.003** | **0.003** |
| 0.35 | **0.096** | **0.018** | **0.005** | **0.003** | **0.003** | **0.003** |
| 0.4 | 0.193 | **0.003** | **0.003** | **0.010** | **0.003** | **0.003** |
| 0.45 | 0.284 | **0.003** | **0.003** | **0.013** | **0.003** | **0.003** |
| 0.5 | 0.198 | **0.023** | **0.003** | **0.068** | **0.003** | **0.003** |
| 0.55 | **0.096** | **0.070** | **0.065** | 0.164 | **0.003** | **0.003** |
| 0.6 | 0.109 | **0.081** | **0.052** | 0.102 | 0.161 | **0.091** |
| 0.65 | **0.060** | **0.036** | **0.036** | **0.003** | **0.078** | **0.073** |
| 0.7 | **0.049** | **0.018** | **0.018** | **0.003** | **0.003** | **0.003** |
| 0.75 | **0.031** | **0.010** | **0.010** | **0.003** | **0.003** | **0.003** |
| 0.8 | **0.029** | **0.005** | **0.008** | **0.003** | **0.003** | **0.003** |
| 0.85 | **0.044** | **0.003** | **0.003** | **0.003** | **0.003** | **0.003** |
| 0.9 | 0.138 | **0.003** | **0.003** | **0.010** | **0.003** | **0.003** |
| 0.95 | 0.516 | **0.016** | **0.016** | **0.010** | **0.003** | **0.003** |

Notes: see Table 7.

### 5.4.2 New evidence from EPU variable

Large literature always sets the U.S. partisan conflict and EPU as the uncertainty factor indicator for the U.S. since there exists a close link between and these two variables. Therefore, in this section, we testify our result robustness by examining whether U.S. EPU has a similar effect on oil and gold prices (returns and volatility) with U.S. partisan conflict. Tables 11 and 12 display the results of causality in quantile tests. Note that the EPU from the U.S. mainly affects the oil returns when the crude oil market is in a bearish state (lower quantiles). By contrast, EPU matters for gold returns only when the gold market is in a bullish scenario (higher quantiles). In addition, for the volatility of oil and gold, the predictability of EPU covers the entire distribution of volatility.

**Table 11** Quantile causality test results (sub-sampling *p*-values) from U.S. EPU to returns of oil and gold prices.

| | Lag | [0.1, 0.9] | 0.1 | 0.2 | 0.3 | 0.4 | 0.5 | 0.6 | 0.7 | 0.8 | 0.9 |
|---|---|---|---|---|---|---|---|---|---|---|---|
| *Oil* | 1 | 0.0266 | **0.0740** | **0.0030** | **0.0355** | **0.0799** | 0.3580 | 0.8757 | 0.0976 | 0.1243 | 0.2485 |
| *returns* | 2 | 0.0207 | **0.0740** | **0.0030** | **0.0562** | 0.1953 | 0.7604 | 0.8787 | 0.1302 | 0.2012 | 0.3846 |
| | 3 | 0.0947 | **0.0710** | **0.0030** | **0.0858** | 0.1805 | 0.8639 | 0.5769 | 0.2337 | 0.3550 | 0.2160 |
| *Gold* | 1 | 0.0148 | **0.0473** | 0.1065 | 0.3817 | 0.4290 | 0.3964 | **0.0030** | **0.0562** | **0.0030** | **0.0503** |
| *returns* | 2 | 0.0296 | **0.0355** | 0.1006 | 0.8757 | 0.7308 | 0.4053 | 0.1893 | **0.0828** | **0.0030** | **0.0030** |

|  | 3 | 0.0207 | **0.0148** | **0.0030** | 0.6331 | 0.5592 | 0.2692 | **0.0592** | **0.0503** | **0.0059** | **0.0059** |

Notes: see Table 7.

**Table 12** Quantile causality test results (sub-sampling *p*-values) from U.S. EPU to price volatility of oil and gold market.

|  | Lag | [0.1, 0.9] | 0.1 | 0.2 | 0.3 | 0.4 | 0.5 | 0.6 | 0.7 | 0.8 | 0.9 |
|---|---|---|---|---|---|---|---|---|---|---|---|
| *Oil* | 1 | **0.0030** | **0.0030** | **0.0030** | 0.2367 | 0.5385 | 0.1420 | **0.0030** | **0.0030** | **0.0030** | 0.0503 |
| *volatility* | 2 | **0.0030** | **0.0030** | **0.0030** | **0.0030** | **0.0030** | **0.0030** | **0.0030** | **0.0030** | **0.0030** | **0.0030** |
|  | 3 | **0.0030** | **0.0030** | **0.0030** | **0.0030** | **0.0030** | **0.0030** | **0.0030** | **0.0030** | **0.0030** | **0.0030** |
| *Gold* | 1 | **0.0030** | 0.0710 | **0.0207** | **0.0266** | 0.0858 | 0.7544 | 0.2426 | **0.0030** | **0.0030** | **0.0237** |
| *Volatility* | 2 | **0.0030** | **0.0030** | **0.0030** | **0.0030** | **0.0030** | **0.0030** | **0.0089** | **0.0030** | **0.0030** | **0.0030** |
|  | 3 | **0.0030** | **0.0030** | **0.0030** | **0.0030** | **0.0030** | **0.0030** | **0.0178** | **0.0030** | **0.0030** | **0.0030** |

Notes: see Table 7.

## 6. Conclusions

In recent years, US politics has been characterized by a high degree of partisan conflict, which has led to increasing polarization and high policy uncertainty. Given the importance of the US in the global commodity market, we employ the novel technique of causality-in-quantiles test to examine the ability of U.S. partisan conflict index in predicting the returns and volatility of oil and gold prices, using monthly data covering the period of January 1981 to October 2017. The main empirical findings are summarized as follows.

First, there is strong evidence in favor of the significant predictable effect of U.S. partisan conflict index on the oil and gold returns at the tails of the conditional distribution of oil and gold returns. Furthermore, it is found that oil returns and gold returns have different responses to US partisan conflict under different market conditions. More specifically, for oil returns, the partisan conflict has a strong predictive power on the oil returns when the crude oil market is in a bearish state (oil returns at the lower quantiles), however, under a bullish state (oil returns at the higher quantiles), the impact of the partisan conflict on oil returns is not significant. By contrast, for gold returns, partisan conflict matters only when the gold market is performing above its normal (average) mode, i.e., in bullish scenarios. It is found that for gold returns at lower quantiles, the partisan conflict index has limited predictive power on the gold returns, with the 0.2 quantiles being an exception. Second, we find that the US partisan conflict index significantly affects the volatility of oil and gold prices over the entire conditional distribution, i.e., at various phases of the oil and gold market. Finally, a robustness exercise using more quantiles to represent the market states and EPU as the uncertainty indicator, the empirical results support the findings.

The results offer some meaningful implications to investors and policymakers. For example, the study shows that the US partisan conflict index can affect the lower quantiles of conditional distribution for oil returns, but is less likely to affect the higher quantiles of oil returns. This indicates that more attention should be drawn to track and monitor the U.S. partisan conflict risk when the oil market is in a bearish state. However, for gold returns, partisan conflict matters for the gold returns only when the gold market is in a bullish scenario. This suggests that gold investors should adopt more prudent investment strategies when the gold market is in a bullish state. The volatility of oil and gold are affected by the partisan conflict in the United States. Therefore, the investors who choose the volatility of oil and gold as the monitoring index should pay close attention to the politics of the United States. Besides, we find significant heterogeneity in the response of two types of strategic commodities, crude oil and gold, to US partisan conflicts. Therefore, for investors, if both oil and gold are included in their asset portfolio basket, they should adopt different risk diversification strategies in the face of impact stemming from conflicts among US political parties.

**Acknowledgements**

We gratefully acknowledge the financial support from the National Natural Science Foundation of China [No. 71431008, 71850012, 71790593], National Social Science Foundation of China [No. 19AZD014], Major special Projects of the Department of Science and Technology of Hunan province [No. 2018GK1020] and a Project Funded by the Applied Economics of Nanjing Audit University of the Priority Academic Program Development of Jiangsu Higher Education Institutions [No. [2018] 87].

**Appendix A. BDS test for the nonlinear feature**

In order to motivate the use of the causality test in quantiles, we investigate the possibility of nonlinearity in the relationship between the U.S. partisan conflict index and returns and volatility of oil and gold prices. To this end, following Balcilar et al. (2017a), we apply the BDS test (Broock et al., 1996) on the residuals of the returns and volatility of oil and gold price equation of the VAR(8) involving (relative) the U.S. partisan conflict, respectively. The BDS test is one of the most popular tests for nonlinearity. It is carried out by testing if increments to a data series are independent and identically distributed (i.i.d.). The test is asymptotically distributed as standard normal under the null hypothesis of i.i.d. increments. The basis of the BDS test is the concept of a correlation integral. A correlation integral is a measure of the frequency with which temporal patterns are repeated in the data.

The results of the BDS test are reported in Table A.1. As shown in panel 1 of Table 6, for the returns and volatility series of oil and gold prices, the null hypothesis of i.i.d. residuals are strongly rejected at 1% level of significance across various dimensions (m). This observation combines results from the BDS test. From the panel 2 and 3, we also see that for the residuals of the returns and volatility of oil and gold price equation of the VAR(8) involving (relative) the US partisan conflict also pass the BDS test at the 1% significance level. It indicates the relationship between the U.S. partisan conflict index and returns and volatility of oil and gold prices is nonlinearity and implies that the Granger causality tests based on a linear framework are likely to suffer from misspecification. In other words, the results of the linear test for Granger non-causality cannot be deemed robust and reliable.

**Table A.1** BDS tests for nonlinear.

| Panel 1: BDS test for each variable | | | | | |
|---|---|---|---|---|---|
| $m$ | 2 | 3 | 4 | 5 | 6 |
| *Oil returns* | 0.041*** | 0.065*** | 0.084*** | 0.095*** | 0.097*** |
| *Gold returns* | 0.014*** | 0.030*** | 0.038*** | 0.046*** | 0.050*** |
| *Oil volatility* | 0.185*** | 0.309*** | 0.390*** | 0.440*** | 0.469*** |
| *Gold volatility* | 0.196*** | 0.331*** | 0.422*** | 0.483*** | 0.522*** |
| Panel 2: BDS test for the residuals of commodity price changes equation of the VAR model with partisan conflict | | | | | |
| $m$ | 2 | 3 | 4 | 5 | 6 |
| *Oil returns* -VAR(8) | 0.022*** | 0.040*** | 0.052*** | 0.060*** | 0.065*** |
| *Gold returns*-VAR(8) | 0.015*** | 0.031*** | 0.040*** | 0.049*** | 0.053*** |
| Panel 3: BDS test for the residuals of commodity price volatility equation of the VAR model with partisan conflict | | | | | |
| *Oil volatility*-VAR(8) | 0.035*** | 0.059*** | 0.074*** | 0.079*** | 0.078*** |
| *Gold volatility*-VAR(8) | 0.019*** | 0.035*** | 0.048*** | 0.056*** | 0.057*** |

Notes: The *** indicates significance at the 1% level. The parameter m is the embedding dimension. The lag parameters are selected based on the AIC by using a VAR model. *m* stands for the embedded dimension.